# A state variable for crumpled thin sheets

Omer Gottesman[1], Jovana Andrejevic[1], Chris H. Rycroft[1], Shmuel M. Rubinstein[*1]

[1] John A. Paulson School of Engineering Applied Sciences, Harvard University,

Cambridge, MA 02138, USA.

[*]Author for correspondence (E-mail: shmuel@seas.harvard.edu)

**Despite the apparent ease with which a sheet of paper is crumpled and tossed away, crumpling dynamics are often considered a paradigm of complexity[1-5]. This complexity arises from the infinite number of configurations a disordered crumpled sheet can take. Here we experimentally show that key aspects of crumpling have a very simple description; the evolution of the damage in crumpling dynamics can largely be described by a single global quantity – the total length of all creases. We follow the evolution of the damage network in repetitively crumpled elastoplastic sheets, and show that the dynamics of this quantity are deterministic, and depend only on the instantaneous state of the crease network and not at all on the crumpling history. We also show that this global quantity captures the crumpling dynamics of a sheet crumpled for the first time. This leads to a remarkable reduction in complexity, allowing a description of a highly disordered system by a single state parameter. Similar strategies may also be useful in analyzing other systems that evolve under geometric and mechanical constraints, from faulting of tectonic plates to the evolution of proteins.**

From collapsed hulls of ships to discarded mathematical theorems written on a white piece of paper, many thin sheets end their life cycle as crumpled heaps. Nevertheless, the dynamics by which an initially flat sheet develops into a disordered and elaborate three-dimensional network of folds are often considered a hallmark example of disordered and complex systems[1,5,6]. When a thin sheet is crumpled, elastic energy focuses to point and line singularities termed d-cones and stretching ridges respectively[1,7-11]. The localization of stresses to individual sharp folds is driven by the local minimization of the elastic deformation energy[8,12-15]. However, the crumpled sheet is never at its global energy minimum and the folds network structure is determined dynamically; as a sheet is confined, existing defects rearrange and new ones are created[16,17]. In elastoplastic sheets, the material scars where the localized stress exceeds the plastic yield threshold[18,19]. In this case, although defects may still migrate, they leave a furrow-like scar in their wake[20]. Consequently, the detailed history of the crumpling dynamics is written into the intricate pattern of creases observed when the sheet is unfolded. No two crumpled sheets are identical.

Geometrical and mechanical constraints forbid the smooth sheet from folding to a ball without accumulating damage in the form of a disordered network of creases. In principle, once the damage is done, the sheet can capitalize on these existing degrees of freedom and fold back smoothly into a ball, without creating new scars. Nevertheless, despite the tendency of the sheet to bend along preexisting scars, it is impossible to crumple the sheet again without creating new folds. Unless the order of folding is reproduced exactly, the system quickly jams at a state in which it cannot conform to further compression; thus, ridges must break[17] and new folds are created. As the process of crumpling and uncrumpling of the sheet is repeated, the network of creases becomes increasingly complex. It is unclear, however, if this chaotic process continues indefinitely or asymptotes to a maximally crumpled state in which the sheet smoothly folds along existing creases and no new scars are created.

We experimentally examine the evolution of the crease patterns in thin sheets that are repeatedly crumpled $n$ times and find that the change of the total length of all creases, $\ell$, is not chaotic at all. Instead, it is a deterministic function whose evolution depends only on its current value. Strikingly, the accumulation of damage does not depend on the crumpling history or the structure of the crease network. Thus, $\ell$ can be interpreted as a state variable of the crumpled sheet. We find that the increase in total crease length, $\delta\ell$, for a given crumpling iteration, $n$, decays exponentially with $\ell$ and that these dynamics are described by a phenomenological equation for the evolution of the damage network. By analyzing this equation in the limit of $n=1$, we precisely resolve the dynamics of initial crumpling a smooth sheet into a ball, a well-studied problem.

The development of damage networks is investigated by repeatedly crumpling elastoplastic thin sheets of Mylar. The mechanical durability of Mylar under repeated crumpling cycles makes it an ideal material for this study: even after accumulating damage over hundreds of repeated crumpling tests, the Mylar sheet retains its spring-like, crackling characteristic. Square sheets, 30 μm thick with side length of $L=100$ mm, are rolled into a cylinder, inserted into a cylindrical container of diameter $D=27$ mm, and compressed uniaxially to a given gap $\Delta < L$, as shown schematically in Fig. 1a. The dimensionless compaction parameter $\widetilde{\Delta} \equiv \Delta/L$ can take values between 0 and 1. After compression the sheets are unfolded and their 3D structure is scanned using a custom laser profilometer[3] as demonstrated in Fig. 1b. To deduce the pattern of damage after every crumpling/unfolding iteration, the measured height profiles of the creases are locally fitted to a surface and converted into a map of mean curvatures. As the crumpling dynamics are dominated by the localization of stresses into folds, the curvature map is dominated by sharp valleys (red) and ridges (blue) with high positive and negative curvatures respectively, as shown in Fig. 1c. Creases are detected by applying the Canny edge detection algorithm to the curvature map. Before measuring the length of creases by summing over pixels determined as edges, the data is cleaned by removing small detected edges below a threshold of a minimal number of connected pixels. Our main results are insensitive to the choice of the parameters of the edge detection algorithm, or to the scale over which the height map is fitted. We then track the evolution of the crease network as a function of the number of crumpling/unfolding iterations, $n$, as shown for a typical example in Fig. 1c and Supplementary Movie 1. The sheets are carefully flattened between every crumpling iteration to replicate the initial conditions of each crumpling as closely as possible.

When the thin sheet is repeatedly crumpled, the damage network evolves as progressively more creases are created. These dynamics lead to a monotonic increase in the total length of both valleys and ridges, $\ell_v$ and $\ell_r$ respectively, as seen for a typical example in Fig. 2a. The rate at which $\ell_v$ and $\ell_r$ increase slows down with $n$, indicating that when more creases are present, the sheet tends to fold along the already existing plastic scars rather than create new folds. Changing $\widetilde{\Delta}$ changes the rate at which creases accumulate, as shown in Fig. 2b for the total length of all creases, $\ell = \ell_r + \ell_v$, for $\widetilde{\Delta}$ ranging from 0.9 to 0.045 and representing data from 507 individual scans. Despite the chaotic nature of crumpling, the evolution of $\ell$ with $n$ is strikingly reproducible. This can be seen, for example, by the open yellow markers in Fig. 2b, which show $\ell(n)$ curves for five different experiments in which sheets are crumpled repetitively to $\widetilde{\Delta} = 0.36$.

As the sheet is re-crumpled, preexisting creases, where the material is weakened, function as mechanical hinges along which the sheet may bend without creating new scars. However, as the compression increases, the sheet often deforms into a jammed configuration in which it cannot compress further by only bending along existing creases. This inevitably leads to the creation of new scars, consistent with the curves shown in Fig. 2b, in which the $\ell(n)$ curves do not plateau. We find that $\ell(n) = a \log(1 + b\,n)$ is a good fit to all $\ell(n)$ curves, as shown for a specific value of $\widetilde{\Delta}$ in the inset to Fig. 2c, where $a$ and $b$ are fitting parameters which depend on $\widetilde{\Delta}$, as shown in Fig. 2c.

For large $\widetilde{\Delta}$ creases accumulate at a slower rate, consistent with the observation that $a$ decreases linearly with $\widetilde{\Delta}$ and $b$ decreases as $\widetilde{\Delta}^{-1}$, as shown in Fig. 2c. Note that since no creases are created when the sheet is not compressed, $a(\widetilde{\Delta} = 1) = 0$, as expected. Thus, $\ell$ and $n$ can be rescaled by $(1 - \widetilde{\Delta})$ and $\widetilde{\Delta}$ respectively, leading to a remarkable collapse of all $\ell(n)$ curves, as shown in Fig. 2d. Moreover, all our data can now be replotted and fitted by

$$\ell(n) = c_1(1 - \widetilde{\Delta}) \log\left(1 + \frac{c_2 n}{\widetilde{\Delta}}\right), \qquad (1)$$

with only two fitting parameters - $c_1 = 5200 \pm 200$, and $c_2 = 0.063 \pm 0.005$ - as shown in the inset to Fig. 2d.

Changing $\widetilde{\Delta}$ varies the rate at which creases accumulate as well as the statistics of the crease pattern, as can be seen by comparing two typical crease patterns with similar $\ell$ shown in Fig. 3a and b. When crumpling a sheet few times to a small $\widetilde{\Delta}$ (Fig. 3a), creases tend to be relatively long and uniformly distributed across the sheet. The same $\ell$ can be obtained by crumpling a sheet many times less vigorously to larger $\widetilde{\Delta}$'s. However, in this case the pattern is dominated by short, more localized creases (Fig. 3b).

So far we only addressed crumpling protocols where one $\widetilde{\Delta}$ was used repeatedly. We test whether the evolution of $\ell(n)$ is history dependent by implementing a new crumpling protocol with two values of $\widetilde{\Delta}$. $\ell_{\widetilde{\Delta}_1,\widetilde{\Delta}_2}(n)$ is measured by initially crumpling a sheet $n_0$ times to $\widetilde{\Delta}_1$, and then crumpling the same sheet several times to a different $\widetilde{\Delta}_2$, as shown in Fig. 3c-d. We compare $\ell_{\widetilde{\Delta}_1,\widetilde{\Delta}_2}(n)$ to the curves obtained by crumpling a fresh sheet repeatedly to a single $\widetilde{\Delta}_i$, $\ell_{\widetilde{\Delta}_i}$, and note that $\ell_{\widetilde{\Delta}_1,\widetilde{\Delta}_2}(n)$ indeed deviates from $\ell_{\widetilde{\Delta}_1}(n)$ for $n > n_0$. Remarkably, $\ell_{\widetilde{\Delta}_1,\widetilde{\Delta}_2}(n > n_0)$ overlaps perfectly with $\ell_{\widetilde{\Delta}_2}(n)$ by a shift along the $n$ axis, marked by open squares in Fig. 3c-d. The overlap of the two curves demonstrates that when two sheets with similar $\ell$ but different crumpling histories are crumpled repetitively to the same $\widetilde{\Delta}$, the evolution of $\ell$ for both is identical. This implies that the evolution of $\ell$ is independent of the crumpling history. Furthermore, as the crumpling history determines the structure of the crease pattern, this history independence also implies that the evolution of the global quantity $\ell$ is independent of the local statistics of the structure; hence, the evolution of $\ell$ is determined solely by its instantaneous value. Drawing an appealing analogy with statistical physics, the history independence of the evolution of $\ell$ suggests that this observable can be thought of as a macroscopic state variable quantifying the "crumpledness" of a damaged sheet. Identifying a global quantity that evolves independently of the details of the pattern significantly reduces the complexity of this system.

Traditionally, crumpling is considered a random and disordered process. The crease pattern obtained for a given sheet is specific to the details of the crumpling dynamics, and is thus impossible to reproduce perfectly. However, as $\ell$ is a state variable with a known functional dependence on $\widetilde{\Delta}$ and $n$, it is a striking corollary that it is possible to fully predict the evolution of $\ell$ for any arbitrary crumpling sequence. $\delta\ell_{\widetilde{\Delta}}$ can be estimated by differentiating Eq. 1 with respect to $n$, yielding

$$\delta\ell_{\widetilde{\Delta}} \approx \frac{\partial \ell}{\partial n} = \frac{c_1 c_2 (1 - \widetilde{\Delta})}{\widetilde{\Delta}} \frac{1}{1 + \frac{c_2 n}{\widetilde{\Delta}}} = \frac{c_1 c_2 (1 - \widetilde{\Delta})}{\widetilde{\Delta}} e^{-\frac{\ell}{c_1(1-\widetilde{\Delta})}}, \qquad (2)$$

where the second equality is equivalent to Eq. 1. The total length of new creases created at any crumpling iteration, $\delta\ell_{\widetilde{\Delta}}$, is a function only of $\widetilde{\Delta}$ and of $\ell$ (measured before the current crumpling cycle) and not a function of $n$. Note that equation (2) is an equation of state for the damage evolution, highlighting that $\ell$ is always memoryless. Through iterative summations of Eq. 2, $\ell(n)$ can be predicted for any crumpling protocol given as a sequence of $\widetilde{\Delta}$'s

We crumple a sheet 12 times according to the protocol described in Fig 4a. (bottom), i.e. we crumple the sheet 6 consecutive times, each time to a smaller volume than the previous crumple (a smaller $\widetilde{\Delta}$), and then repeat this cycle. The measured $\ell(n)$ curve for this crumpling protocol is in excellent agreement with the prediction obtained from Eq. 2, as seen in Fig. 4a (top). Note that there are no fitting parameters used in Fig. 4a, as $c_1$ and $c_2$ are extracted from the data shown in Fig. 2c. This is possible only because $\ell(n)$ is a state variable that evolves without memory.

The agreement between experiment and prediction demonstrated in Fig. 4a further indicates that the evolution of $\ell$ is history independent, and supports the claim that it may be treated as a state variable of crumpling. Such a statement will be more meaningful if it could be applied to a general loading configuration; however, so far we have only considered this history independence for fold networks created by uniaxially crumpling sheets in a cylindrical configuration. We therefore repeat the crumpling protocol shown in Fig. 4a and measure $\ell(n)$ for two pre-creased sheets with distinctively different loading histories; one sheet was first hand crumpled into a ball, and the other folded along straight lines to create the initial creases pattern shown in the insets to Fig 4b. For both sheets, Eq. 2 with $c_1$ and $c_2$ extracted from Fig. 2c still accurately predicts the evolution of $\ell$, as shown in Fig 4b.

The history independence of $\ell$ for sheets that are repeatedly crumpled allows us to take an unconventional approach to understanding the dynamics of crumpling. By this approach we gain direct insight into the crumpling process, resolving the question regarding the evolution of the crease network as a smooth sheet is confined to an increasingly shrinking volume. That is, how does $\ell$ depend on $\widetilde{\Delta}$? Rather than tracking the evolution of $\ell(n)$ as a sheet is repeatedly crumpled to a given $\widetilde{\Delta}$, we now inspect $\ell(\widetilde{\Delta})$ for a given $n$. In Fig. 4c we re-plot the data presented in Fig. 2a as the surface $\ell(n, \widetilde{\Delta})$, where we highlight two $\ell(\widetilde{\Delta})$ traces. The history independence of the crumpling dynamics corresponds to path independence along the $(n, \widetilde{\Delta})$ phase space; thus, $\ell(\widetilde{\Delta})$ curves at constant $n$ are mathematically and physically meaningful even though it requires many different sheets, with their unique network of creases, to experimentally generate one such curve.

For large $n$, $\ell(n)$ curves are smooth and $\ell(n = const, \widetilde{\Delta})$ curves are traced well by Eq. 1, as seen in the inset to Fig. 4d. By examining Eq. 1 at $n=1$ we can obtain a prediction for the accumulation of creases as a smooth sheet is crumpled for the first time. The great advantage of this approach is that the phenomenological Eq. 1 and the two constants $c_1$ and $c_2$ are obtained by fitting the entire $\ell(n, \widetilde{\Delta})$ surface - a fit dominated by the large $n$ data, where the noise is significantly reduced. As $c_2 = 0.063$, for $n = 1$, Eq.1 is approximated by:

$$\ell(n = 1, \widetilde{\Delta}) \equiv \ell_1(\widetilde{\Delta}) = \frac{c_1 c_2 (1 - \widetilde{\Delta})}{\widetilde{\Delta}}, \qquad (3)$$

Eq.3 is in striking agreement with the experimental results for $n=1$, as seen in Fig. 4d.

Eq.3 indicates that for small compression ($\widetilde{\Delta} \approx 1$), $\ell_1$ is proportional to the strain applied by the piston, $(1 - \widetilde{\Delta})$, while for very large compression ($\widetilde{\Delta} \to 0$), $\ell_1$ is inversely proportional to $\widetilde{\Delta}$.

A heuristic 1D model of crumpling, described schematically in the insets to Fig. 4d, provides intuition for the behavior of $\ell(n = 1, \widetilde{\Delta})$ in the two limits. For small compression, the sheet can be compacted by creating a circumferential fold along the sheet, which at this stage of the crumpling is roughly cylindrical. This fold serves as a hinge that the cylinder can bend along without the need to create new folds. The sheet can thus compress continuously, creating new hinges when the freedom of travel provided by the existing hinges runs out, as described in inset schematic 1. In this regime $\ell$ is proportional to the number of hinges created, leading to a linear relation between $\ell$ and the distance the piston compressed the sheet, i.e. $\ell \sim (1 - \widetilde{\Delta})$. For large compression, the sheet is tightly packed and all facets of the sheet must be broken when compression is increased. This breaking of all facets doubles the total length of creases for every halving of $\widetilde{\Delta}$, leading to the observed scaling of $\ell \sim \widetilde{\Delta}^{-1}$, and represented in inset schematic 2.

The crumpled state can be thought of as one point in a hyper-dimensional configuration space, where the angle of each individual fold corresponds to a dimension, and the folding dynamics are represented by a trajectory in configuration space. The seemingly unbounded increase of $\ell$ for large $n$ implies that most of these trajectories lead to a dead-end where the system jams; the sheet cannot smoothly compact by deforming along existing creases, instead, to reach the designated compaction new, energetically expensive creases are created. Because of the complexity of such configuration space, it is nontrivial that much of this system's evolution can be captured by a single state equation. In contrast to classical glassy systems, where the dynamics are illustrated as a stroll in a complex energy landscape, our system holds richer dynamics. When our system jams in a local minimum, it supports further confinement by introducing new ridges or valleys, increasing the dimensionality of the system. Furthermore, the energy landscape changes as a result of the interactions between the new and existing folds. It would be meaningful in the future to explore the dynamics of how the energy landscape evolves as folds accumulate. It may also be noteworthy to look for state variables in other systems that evolve under geometric and mechanical constraints. For example, earthquake fault networks which evolve via the accumulation and release of tectonic stresses by the formation of new faults or slip along pre-existing faults that are themselves remnants of its seismic history[21]. More provocatively, we may consider the evolution of functional materials, such as proteins[22-24], where several recent works suggest that through continuous structural alterations, resulting from cyclic loading, genetic complexity is reduced via evolutionary selection to perform a specific mechanical task.

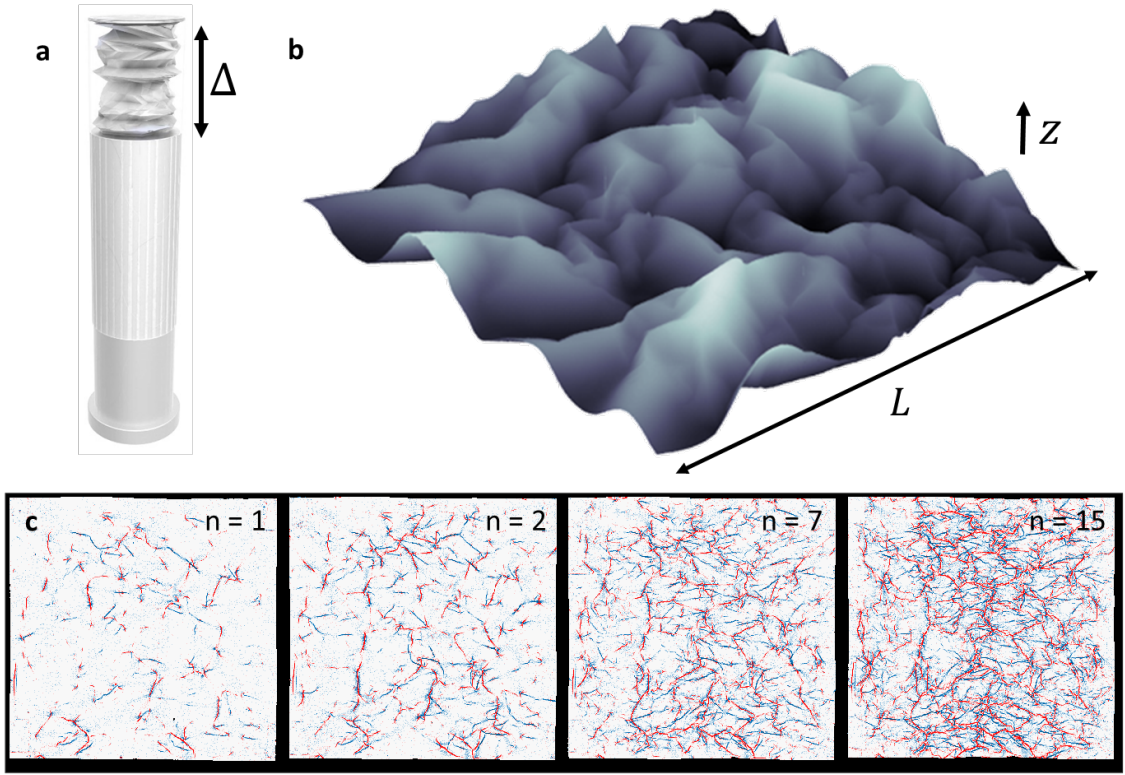

**Figure 1: Crumpled thin sheets scared by ridges and valleys.** (a) Mylar sheets are crumpled uniaxially in a cylindrical container to a given gap, $\Delta$. (b) 3D topography of an unfolded crumpled sheet. Height maps are obtained with a laser profilometer similar to the one designed by Blair and Kudrolli[3]. (c) Mean curvature maps for the scanned surfaces of crumpled thin sheets for n=1, 2, 7 and 15 with $\widetilde{\Delta} = \frac{\Delta}{L} = 0.27$. Red and blue correspond to positive and negative mean curvatures respectively.

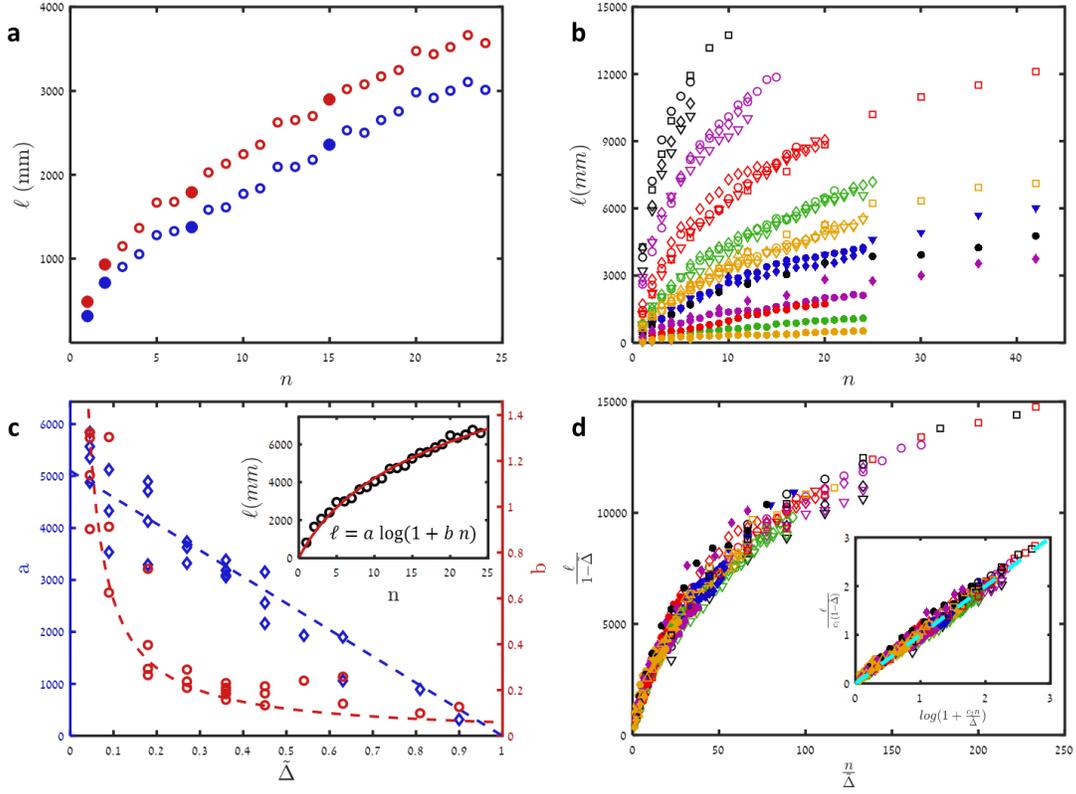

**Figure 2: The evolution of damage networks.** (a) $\ell_r(n)$ for ridges (blue) and $\ell_v(n)$ for valleys (red). Full circles correspond to the four crease patterns shown in Fig. 1 (c). (b) $\ell(n) = \ell_r + \ell_v$ for $\widetilde{\Delta}$ ranging from 0.9 to 0.045 for 507 scans. Different markers of identical colors correspond to different sheets crumpled to the same values of $\widetilde{\Delta}$. (c) . $a$ (blue diamonds) and $b$ (red circles) as a function of $\widetilde{\Delta}$. The dashed lines are the respective fits to $a = c_1(1 - \widetilde{\Delta})$ ($c_1 = 5200 \pm 200$) and $b = c_2/\widetilde{\Delta}$ ($c_2 = 0.063 \pm 0.005$). Note that for $\widetilde{\Delta} \to 0$, $\ell$ diverges for any finite $n$. (inset) Fit (red curve) of a single $\ell(n)$ curve to $\ell(n) = a \log(1 + b n)$. (d) $\ell/(1 - \widetilde{\Delta})$ as a function of $n/\widetilde{\Delta}$ for all data shown in Fig. 2b. collapse, demonstrating that all $\ell(n)$ curves follow Eq. 1 (inset).

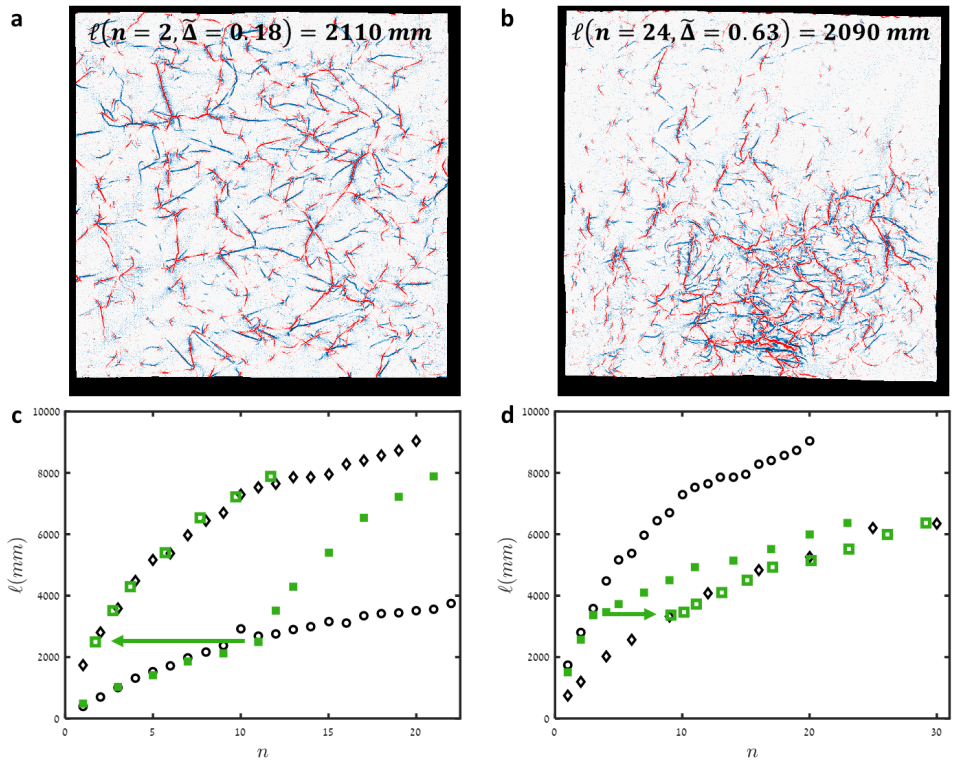

**Figure 3: Accumulation of damage is not history dependent.** (a-b) Scanned surfaces with almost identical $\ell$ created with two different crumpling protocols (c) $\ell(n)$ curves for $\widetilde{\Delta}_1 = 0.45$ and $\widetilde{\Delta}_2 = 0.18$ (black circles and diamonds respectively), and $\ell(n)$ curve for a sheet crumpled $n_0 = 11$ times to $\widetilde{\Delta}_1$ and subsequently crumpled to $\widetilde{\Delta}_2$ (full green squares). The open green squares correspond to shifting of the $n > n_0$ part of the $\ell(n)$ data to the $\widetilde{\Delta}_2$ curve. (d) Same as (c) for $\widetilde{\Delta}_2 > \widetilde{\Delta}_1$ with $\widetilde{\Delta}_1 = 0.18$, $\widetilde{\Delta}_2 = 0.36$ and $n_0 = 3$.

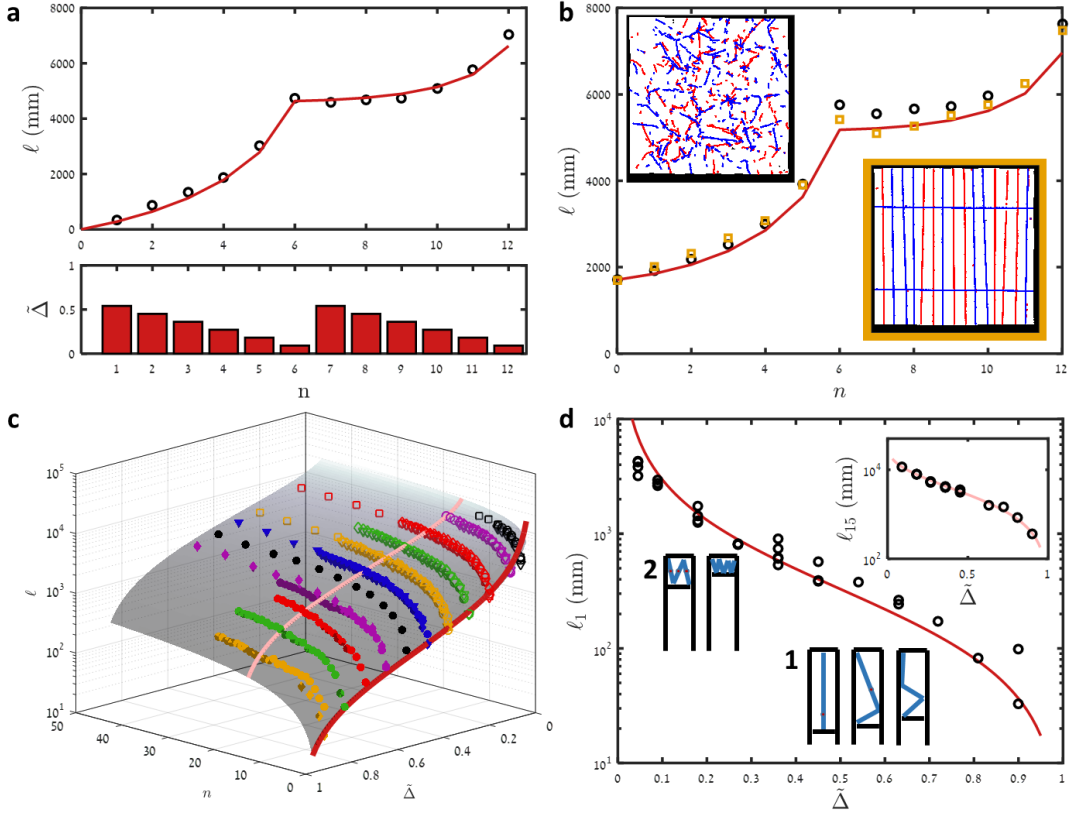

**Figure 4: Crumpling dynamics.** (a) (top) $\ell(n)$ (circles) and prediction based on Eq. 2 (solid red line) for a crumpling protocol given as a sequence of $\widetilde{\Delta}(n)$ (bottom). (b) $\ell(n)$ for sheets initially crumpled by hand (black circles and top left inset) or folded in straight lines (yellow squares and bottom right inset) and then subjected to the same crumpling protocol as in Fig. 4a. Solid red line again shows the prediction based on Eq. 2. Curvature maps are thresholded for clarity. (c) The surface $\ell(n, \widetilde{\Delta})$ given by Eq.1. The pink and red curves correspond to $\ell(n = 15, \widetilde{\Delta})$ and $\ell(n = 1, \widetilde{\Delta})$ respectively. (d) $\ell(n = 1, \widetilde{\Delta})$ and $\ell(n = 15, \widetilde{\Delta})$ (inset) with fit based on Eq. 3 and 1, respectively. A 1D heuristic model predicts $\ell$ to depend linearly on $(1 - \widetilde{\Delta})$ for mild compression ($\widetilde{\Delta} \approx 1$) (schematic 1) and to scale with $\widetilde{\Delta}^{-1}$ for large compression ($\widetilde{\Delta} \approx 0$) (schematic 2).

## Supplementary Discussion

In the main text, $\ell$ is evaluated for each curvature map by applying Canny edge detection, which defines a skeleton of the crease network, and totaling the length of all segments. In the review process, a concern was raised as to how the sensitivity of our measurement depends on crease density. Specifically, could the functional dependence of $\ell$ result from the analysis missing more creases in denser regions? To address this possible vulnerability, here we demonstrate that the scaling of $\ell$ is robust to the choice of detection method by presenting an alternative technique to Canny edge detection. We will refer to the new approach as the *Radon transform method*, as it repurposes the principle behind a Radon transform to augment the signature of creases and reduce noise. We begin by detailing the algorithm of the Radon transform method, demonstrating the consistency of the scaling analysis across the two crease-detection techniques, and lastly validating the robustness of crease detection to varying crease density.

In the Radon transform method, the crease content is mapped to a signal array constructed by integrating the mean curvature along directed paths; this is calculated for each pixel. A strong signal is recovered if an integration path closely coincides with the direction of an extended structure such as a crease; a weak signal is produced by features that are point-like or isotropic, common attributes of noise and fine texture in the data. In detail, a linear integration path is centered about a given pixel, traversing a fixed circular local window. The average curvature along a particular direction is computed by exact numerical integration of the bicubic interpolant on the grid defined by pixel centers. The integration direction is systematically rotated about the central pixel; the maximum average curvature over all path orientations is selected as the signal. This process is repeated for all pixels in the curvature map, resulting in a signal array of only the average curvatures that are a maximum along local, linear paths.

Next, global and local thresholds are applied to the signal array to separate the real creases from the background noise. A global threshold is more permissive of noise but acts uniformly across the signal array; the local threshold considers the neighborhood of each pixel, labeling as creases only those pixels whose intensity is above a set fraction of the maximum signal within their neighborhood. Local thresholding accommodates variations in signal intensity, and thus provides sensitivity to softer (less sharp) creases. After thresholding, the resulting binary array is thinned to produce a skeleton of the crease network. Similar variants of thresholding and thinning are likewise present in the Canny edge detection algorithm.

The key difference between our two approaches is the nature of the signal we used to separate creases from background noise. Canny edge detection identifies peaks in the local gradient of the curvature map as creases; the Radon transform method in turn selects for peaks in local integrals, and thus relies on a fundamentally distinct definition of a crease. There are however some differences during pre-processing; a Gaussian smoothing filter is applied in Canny edge detection, while in the Radon transform method, curvature maps are downsampled for computational efficiency, which likewise has the effect of smoothing. Finally, in the Radon transform method we introduce an additional post-processing step; due to variations in the thickness of the identified crease lines before thinning, the network skeleton may contain artifacts in the form of short segments stemming from a principal crease line. These are cleaned by identifying all branchpoints along a crease, and removing branches whose size in pixels falls below a fixed value. Connected components smaller than the minimum size are also removed. However, we demonstrate in Figure S3 that excluding this post-processing step does not affect the scaling of $\ell$, as such artifacts make up only a small portion of the crease network. The logarithmic dependence of $\ell$ is likewise recovered when the downsampling step is omitted. A progression from the initial curvature map to the final crease network produced by the Radon transform method is illustrated for a typical data set in Figure S1. A downsampling factor of four was found to preserve the integrity of the crease pattern while providing useful speedup in computation. Integrals along 17 equally spaced path orientations on the interval of 0° to 180° were considered at each pixel and the maximum was selected as the signal. We examined a range of integration path lengths up to 8 $mm$, as the integration window defines a new length scale that must accommodate features of varied sizes. While smaller integration paths can detect finer details particularly at low crease densities, they sacrifice some of the advantage afforded by longer paths in accruing a strong signal that is well separated from noise. An integration path length of 3.2 $mm$ suitably mediated such effects and provided a clear crease network as shown. When applied to the complete dataset, the Radon transform method demonstrates clear consistency in the measure of $\ell$ and reproduces the scaling identicaly to Figure 2 of the main text, as shown in Figure S2.

We conclude by evaluating the performance of our Radon transform method for an increasing density of creases. A simple and informative approach is to generate synthetic, high density data by the superposition of lower density crease patterns. Our strategy is as follows: If it can be assumed that the Radon transform method reliably recovers the crease network at low densities, then $\ell$ for a sum of low density patterns can be straightforwardly computed by adding the known lengths of the constituent networks, and subtracting their overlap. This result may then be compared to $\ell$ measured by directly analyzing the synthetic dense patterns. Such an analysis enables us to discern any systematic bias or weakening in our detection ability of creases at higher density. In Figure S4, the curvature maps of thin sheets crumpled to $\tilde{\Delta}= 0.36$ and $\tilde{\Delta}= 0.45$ are added together at corresponding crumpling iterations. For the densest synthetic pattern, $\ell$ is comparable to the highest crease content observed in experiment. A real reference experiment of $\tilde{\Delta}= 0.18$, whose $\ell$ values are close to the calculated crease lengths of the superimposed patterns, is used to visually evaluate the authenticity of the synthetic networks. The simple addition of two patterns can result in cancellations that may both weaken creases and smooth flat regions of the curvature maps; however, by visual comparison with the constituent patterns, these effects do not appear to be prominent, as shown by the resemblance between the synthetic crease patterns and the experimental data in Figure S4. Enforcing bounds on the extremal curvature values of the synthetic maps to correct for sharp, overlapping creases added constructively likewise has negligible effect on the detected creases. As shown in Figure S5, the detected total crease length is in good agreement with the expected result across varied integration path lengths of the Radon transform method. Taken as a whole this analysis suggests that it is very unlikely that the logarithmic dependence of $\ell$ and the remarkable correspondence to the phenomenological equation of state presented in the main text is an artifact of the detection method at higher densities.

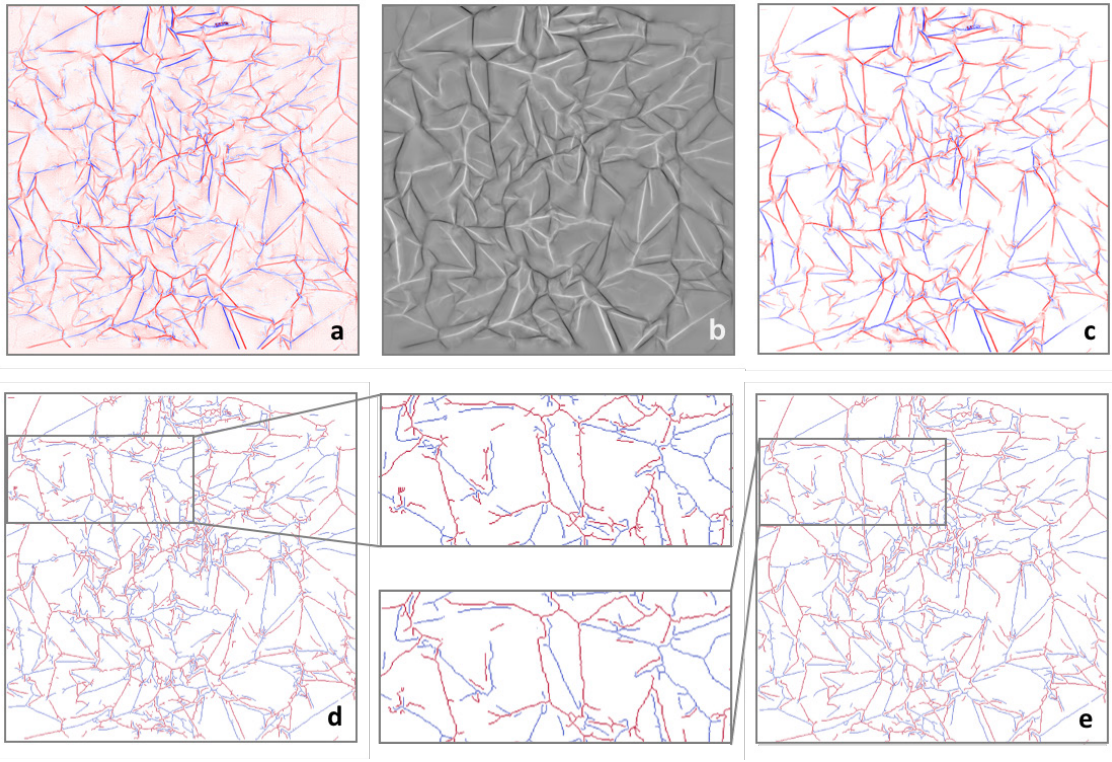

**Figure S1: The Radon Transform Method.** (a) Typical map of mean curvature. (b) A corresponding constructed signal array of maximum average curvature along linear paths local to each pixel. (c) Creases identified after local and global thresholding. (d) Skeleton of the crease network produced by thinning. (e) Final crease network used to compute $\ell$. Insets in (d) and (e) show close-up views of creases before and after the removal of small, spurious segments which arise during thinning due to varying crease thickness.

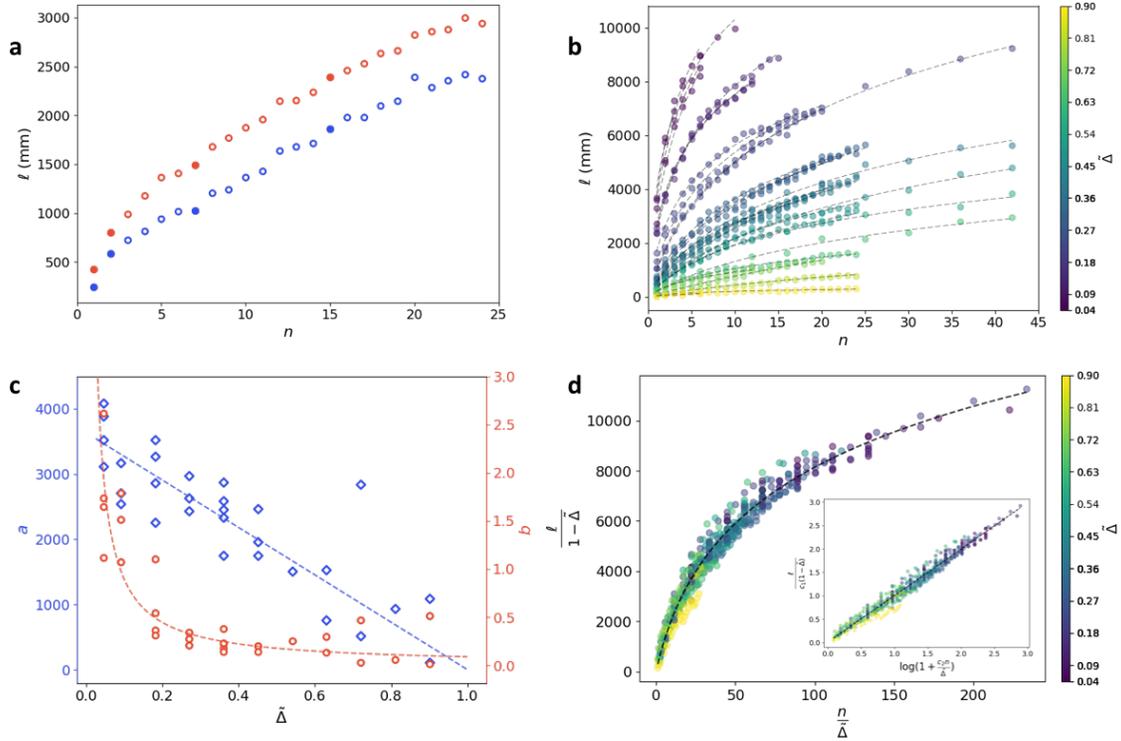

**Figure S2: Reproduced evolution of damage networks.** Figure 2 of the main text is reproduced using the Radon transform method in place of Canny edge detection. Curvature maps labeled using the Radon transform method are downsampled for computational efficiency; thus, measurements of $\ell$ have been rescaled by the downsampling factor to obtain the best quantitative correspondence with the full resolution results. A collapse of all the curves in (b) onto the form of Eq. 1 is clearly recovered and shown in (d), affirming the robustness of the scaling analysis to different crease detection techniques.

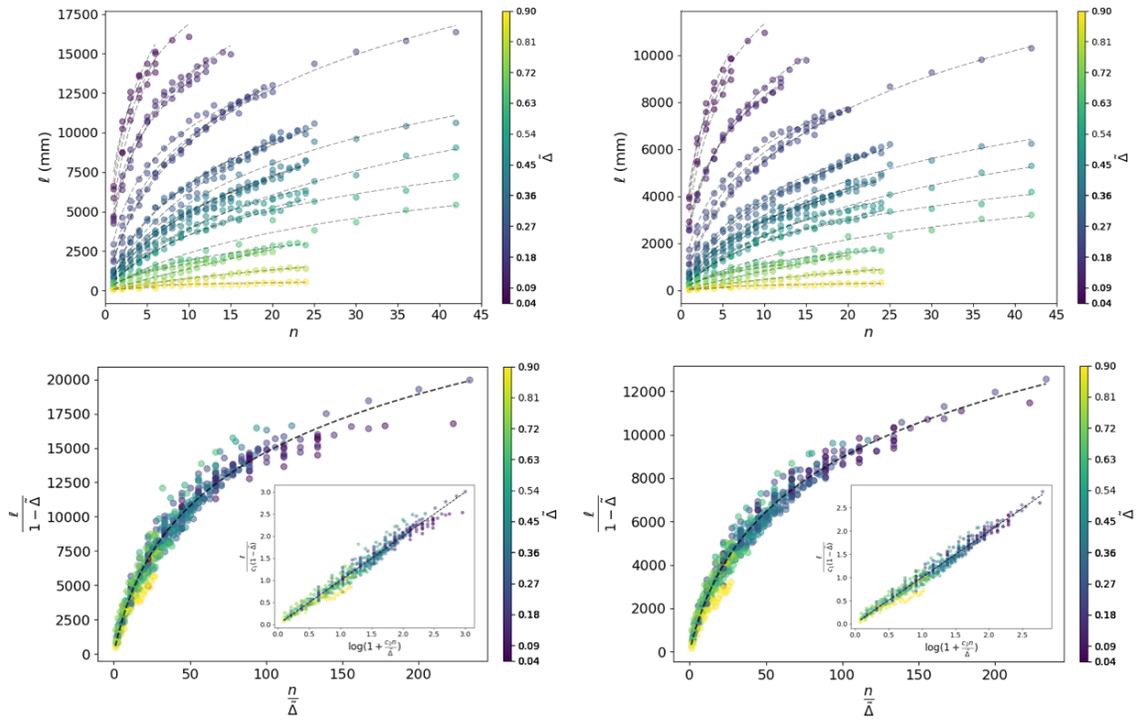

**Figure S3: Robustness of detection to modifications in processing.** Prior to crease detection, curvature maps analyzed using the Radon transform method are downsampled for computational efficiency. After the crease network has been extracted, small artifacts that arise due to thinning are further identified and removed. In this figure, the left column shows the scaling analysis performed without downsampling the curvature maps and without cleaning small artifacts in post-processing. The right column shows the results of including downsampling, but without the artifact removal.

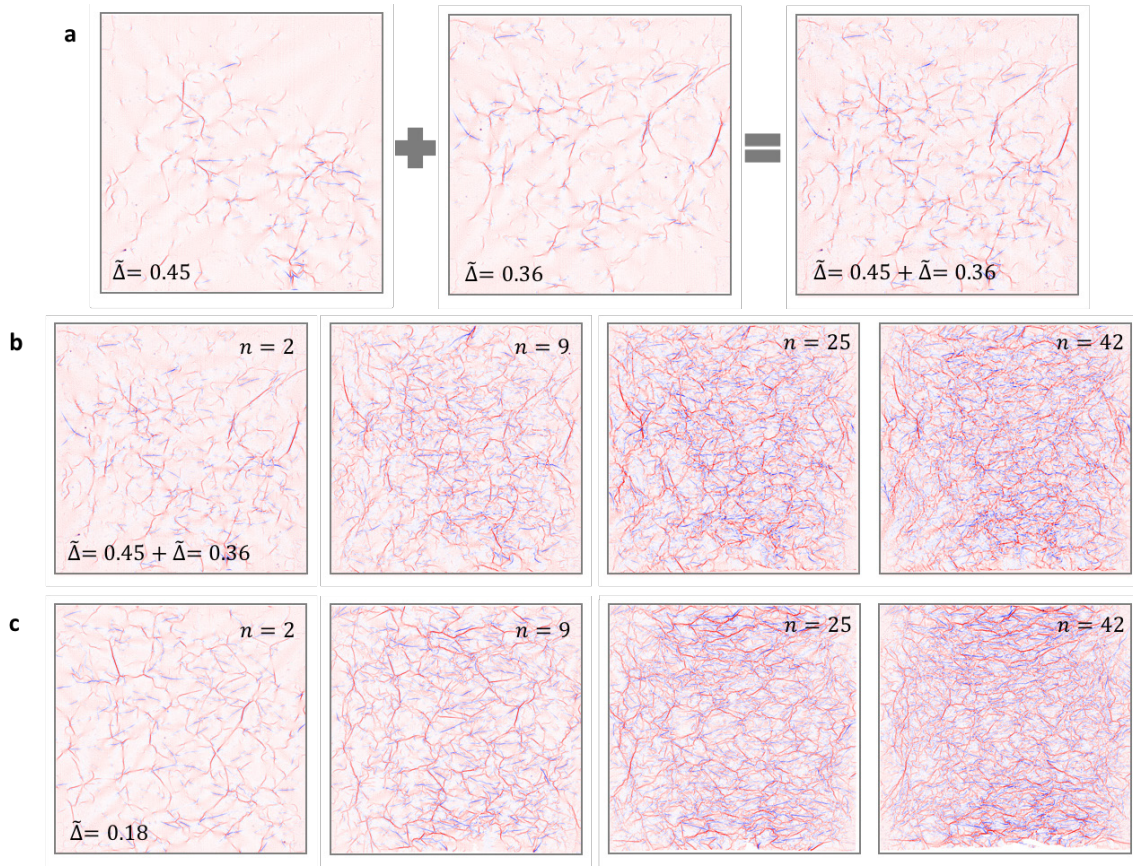

**Figure S4: Generating synthetic high-density crease patterns.** (a) A synthetic, dense crease pattern is generated by numerically summing the curvature maps of two lower density patterns. (b) The curvature maps of sheets crumpled to $\tilde{\Delta}= 0.36$ and $\tilde{\Delta}= 0.45$ are added together at corresponding crumpling iterations, with the resulting synthetic patterns shown for $n = 2, 9, 25$ and $42$. (c) A true experiment crumpled to $\tilde{\Delta}= 0.18$ exhibits a density of damage comparable to the synthetic patterns at corresponding crumpling iterations.

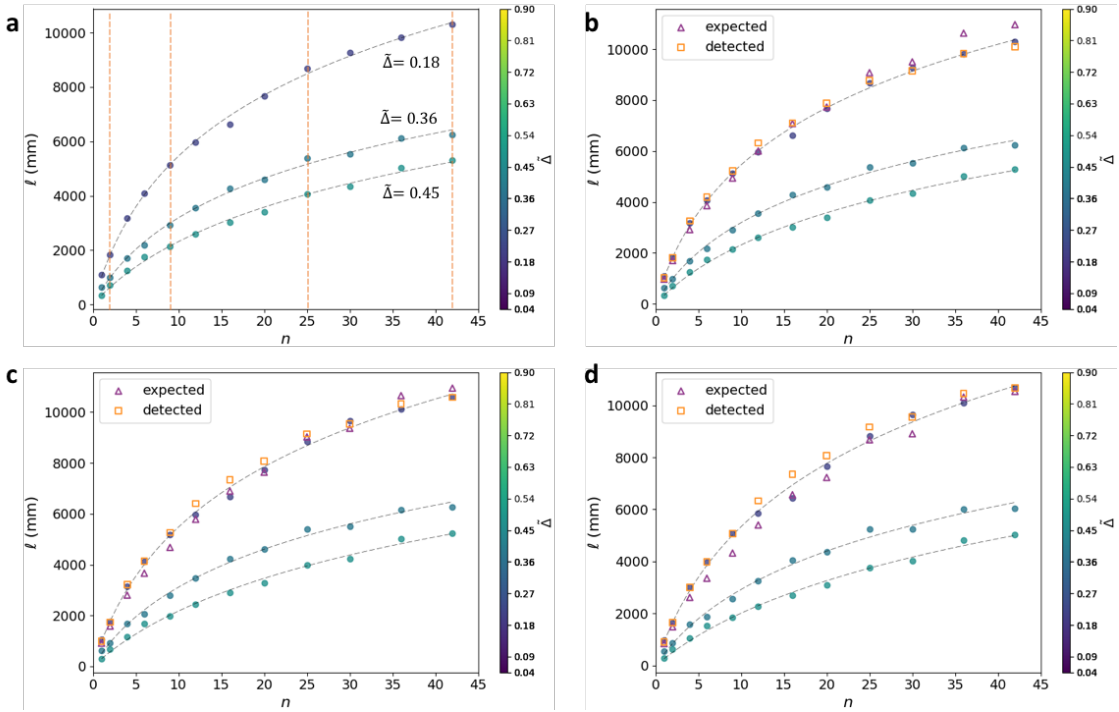

**Figure S5: Testing crease detection for synthetic high-density patterns.** (a) $\ell$ as a function of $n$ for the two superimposed low-density experiments, crumpled to $\tilde{\Delta}= 0.36$ and $\tilde{\Delta}= 0.45$, and the reference high density experiment of $\tilde{\Delta}= 0.18$, from Figure S4. Orange dashed lines correspond to the iterations at which the crease patterns of Figure S4 are produced. (b) $\ell$ vs $n$ for patterns obtained by summing the lengths of the constituent crease patterns, minus their overlap (purple triangles), compared alongside $\ell$ measured by applying the Radon transform method to the added synthetic crease pattern (orange squares). Results correspond to an integration path length of $3.2\ mm$ in the Radon transform method, and the detected crease length, $\ell$, shows good agreement with the expected. (c) and (d) present the same results as (b) for integration path lengths of $4.8\ mm$ and $6.4\ mm$, respectively.